\documentclass[runningheads]{llncs}
\usepackage[T1]{fontenc}
\usepackage{graphicx}
\usepackage{xcolor}
\usepackage{subfigure}
\begin{document}
%
\title{Automatic Voice Classification Of Autistic Subjects\thanks{This publication  is part of the project NODES which has received funding from the MUR – M4C2 1.5 of PNRR funded by the European Union - NextGenerationEU (Grant agreement no. ECS00000036). The work was partially supported by Fondazione TIM under the Italian national project VOCE, call for proposals “Liberi di Comunicare”.}}

%
%
\author{Jessica Vacca\inst{1} \and
Natascia Brondino\inst{2}\orcidID{0000-0002-3128-1592}\and
Fabio Dell'Acqua\inst{3}\orcidID{0000-0002-0044-2998} \and
Anna Vizziello\inst{3}\orcidID{0000-0002-6378-141X}\and
Pietro Savazzi\inst{3}\orcidID{0000-0003-0692-8566}
}
\authorrunning{J. Vacca et al.}
%
\institute{Department of Electrical, Computer, Biomedical Engineering, University of Pavia, Pavia, Italy \\
\email{jesva97.jv@gmail.com}\\
\and
Department of Brain and Behavioral Sciences, University of Pavia, Pavia, Italy\\
\email{givenname.surname@unipv.it}\\
\and
Department of Electrical, Computer, Biomedical Engineering, University of Pavia, \& CNIT Consorzio Nazionale
Interuniversitario per le Telecomunicazioni, Pavia, Italy\\
\email{givenname.surname@unipv.it}\\
}
\maketitle              
\begin{abstract}
Autism Spectrum Disorders (ASD) describe a heterogeneous set of conditions classified as neurodevelopmental disorders.
Although the mechanisms underlying ASD are not yet fully understood, more recent literature focused on multiple genetics and/or environmental risk factors. Heterogeneity of symptoms, especially in milder forms of this condition, could be a challenge for the clinician. In this work, an automatic speech classification algorithm is proposed to characterize the prosodic elements that best distinguish autism, to support the traditional diagnosis. The performance of the proposed algorithm is evaluted by testing the classification algorithms on a dataset composed of recorded speeches, collected among both autustic and non autistic subjects.
\keywords{Speech Recognition  \and Machine Learning \and Signal Classification \and Autism.}
\end{abstract}

\section{Introduction}
Autism spectrum disorders (ASD) consist of a heterogeneous set of neurodevelopmental conditions characterized, at varying levels of complexity and severity, by persistent difficulties in verbal and non-verbal communication and in social interaction, and by patterns of repetitive behaviors and restricted interests \cite{DSM5}. ASD subjects represent a clinical condition with different levels of severity and whose features can be extremely heterogeneous \cite{Volkmar2019}.

The diagnosis of ASD is substantially clinical and it is formulated both in  children and adults by physicians with significant expertise in ASD according to international standardized criteria and supported by gold standard instruments (ADOS, ADI-R). Unfortunately, to date, there are no clear diagnostic biomarkers to support the clinician during the diagnostic process \cite{Hodges2020}. 

From their earliest characterizations \cite{Kanner}, ASDs have been associated with particular speech tones and prosody disorders. Although 70-80\% of individuals with ASD develop functional spoken language, at least half of the population with ASD develops atypical vocal patterns \cite{Fusaroli2017}. Specifically, verbal children with ASD often show some specific acoustic patterns \cite{Mohanta2020}: prosodic features such as monotone tone, reduced stress, flat intonation, and even differences in the harmonic structure of their speech are among the first signs of the disorder \cite{Asgari2021}.

To discuss the matter, it is useful to understand the mechanism behind speech production \cite{Quatieri2001}. The organs of the vocal production system are the lungs (subglottal system), the larynx, and the vocal tract (supraglottal system). Lungs supply energy to the larynx in the form of an airflow that is modulated by the vocal cords. Vibration of the vocal cords turns the airflow into an almost periodic pressure variation or into a noisy sound depending on the action of the cords. This is used to excite the vocal tract system. This latter consists of oral, nasal and pharyngeal resonant cavities that further shape the signal spectrum of modulated airflow. The resulting signal is radiated through the lips. The vibration patterns of the vocal cords and the shape of the vocal tract system can produce different types of sounds. Therefore, the vocal signal generation system can be seen as consisting of an input excitation source to the vocal tract filters. This model is used in the source-filter theory of vocal production, which is based on the assumption that vocal outputs can be analyzed as the response of a series of vocal tract filters \cite{Quatieri2001}.

Specifically, the excitation signal for the vocal tract filter can be classified into one of
 the following list of options:
\begin{itemize}
    \item Periodic glottal vibrations: quasi-periodic signal consisting of variable airflow cycles due to vocal cord vibration;
    \item Noise: source of aperiodic excitation generated when air flows rapidly through an open, non vibrating glottis (suction) or when air flows rapidly through a tight supra-laryngeal constriction (friction). 
    \item Burst or pulse: short pulse of excitation caused by a rapid change in oral atmospheric pressure.
\end{itemize}
The input-output relationship of the vocal tract filter, where the input is the glottal airflow velocity and the output the airflow velocity through the vocal tract itself, can be approximated by a linear resonant filter. The resonant frequencies of the vocal tract are called formant frequencies or simply formants. The main objective of this work is to define appropriate algorithms for automatic extraction and classification of these speech features in order to correctly identify ASD subjects from the analysis of their recorded speech, partly building on previous results \cite{Mohanta2022}.

The rest of this work is organized as in the following: section \ref{FECA} shows the implemented algorithms from extraction of the speech features to subject classification. Section \ref{ER} is devoted to experimental results analysis, while in section \ref{CC}, we comment the obtained results and future research directions. 

\section{Feature Extraction and Classification Algorithms}
\label{FECA}
\begin{figure}
\centering
	{\includegraphics[width=1\textwidth]{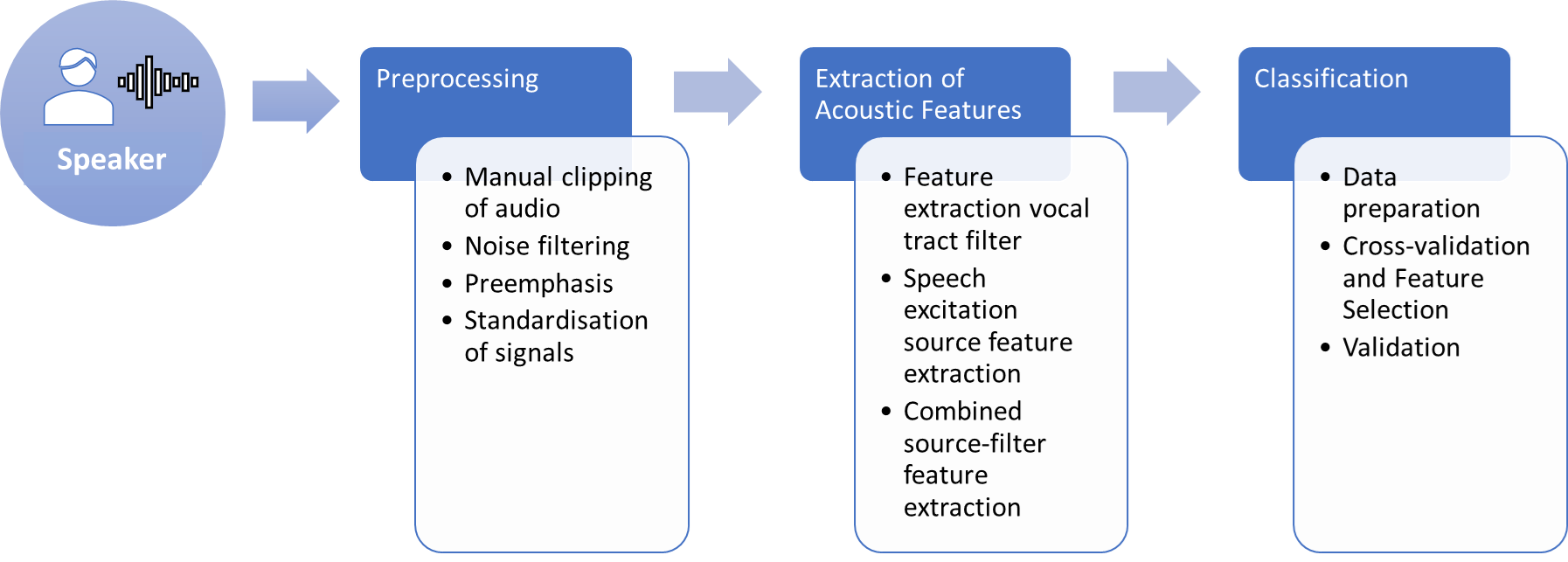}}
	\caption{Pipeline of the implemented algorithm}
	 \label{fig:algorithm}
\end{figure}

In this section, we describe the signal processing chain designed to extract the vocal features of interest in order to correctly classify autistic subjects. Unless otherwise stated, all signal processing algorithms described here were implemented in Matlab \cite{Matlab}. The proposed algorithms were applied to the dataset \cite{ASDBank}, as detailed in the subsequent section \ref{ER}. Information on this dataset can be found in \cite{Hendriks2014,Kuijper2015}.
Fig. \ref{fig:algorithm} provides an overview of the implemented algorithm.

\subsection{Preprocessing and Feature Extraction}
Preprocessing were performed on each audio sequence, and include manual clipping of the signal, noise filtering, pre-emphasis, and signal normalization. 

Manual clipping and noise filtering were done using the open source software Audacity \cite{Audacity}, to remove the silent beginning and ending parts of each audio sequence and filter ambient noise. Regarding this latter, a part of the sequence containing only noise was selected, with a duration sufficient to guarantee good performance; for example, 0.05 s for an audio sequence sampled at 44.1 KHz. The noisy sample was selected to include different types of environmental noise, such as book page turning, and close-range buzzing of the microphone used for recording.

Once the noise profile had been analysed by the software, noise is reduced on the entire sequence of the audio waveform by setting suitable noise reduction parameters. Selection of parameter values was performed manually to avoid distortion of the speech sequences. The parameters include:
\begin{itemize}
    \item Noise reduction set equal to 6 dB.
    \item Sensitivity: Audacity offers a range between 0 and 24 for this parameter. Higher values remove more noise, but with increasing likelihood of suppressing also a fraction of the useful speech signal. Lower values may however produce artefacts in the processed audio signal. We experimentally found the best tradeoff to be 6.
    \item Frequency damping: on a range between 1 and 6. The frequency damping operation consists of reducing the amplitude of selected frequency bands, in order to selectively filter noise in specific frequency ranges where it is thought to prevail. The width of each band is automatically selected according to the damping frequency of the audio sequence. It can be useful for reducing the effects of artifacts that may appear to the the sensitivity parameter setting. The best compromise for this parameter was found to be 6, implying that selective suppression was very effective in noise reduction.
\end{itemize}



The subsequent preprocessing steps were implemented in Matlab \cite{Matlab}.
Pre-emphasis was obtained through a high-pass finite impulse response (FIR) filter to increase the power of the high frequencies of the vocal signal while the low frequencies remain unaffected \cite{Ai2012}:

\begin{equation}
\label{eq_1}
   H(z) = 1 - \alpha z^{-1}
\end{equation}

where $0.9 \le \alpha \le 1$ is the pre-emphasis coefficient that fixes the cut off frequencies of the filter;  it was set to 0.98.

Since the FIR filter (\ref{eq_1}) changes the distribution of energy among frequencies along with the overall energy level, which may impact on the energy-related acoustic characteristics \cite{Vergin1995}, a normalization is applied to allow the comparison among speech signals independently from their amplitude variations. The $i$-th normalized sampled is defined as $S_{Ni}=(S_i-\mu)/\sigma$, where $S_i$ is the $i$-th sample of the vocal signal, $\mu$ is the mean and $\sigma$ is the standard deviation of the whole set of samples.

In order to extract the signal features used by the classification algorithms, the speech signal is here divided in vocal tract and vocal source to better discriminate the differences between the audio sequences of the autistic subjects and the ones recorded from neurotypical individuals. In the following, the employed techniques are described to extract the features of the vocal tract filter, of the excitation source, and of the combined source-filter \cite{Mohanta2020}.

\begin{figure}
\centering
	{\includegraphics[width=1\textwidth]{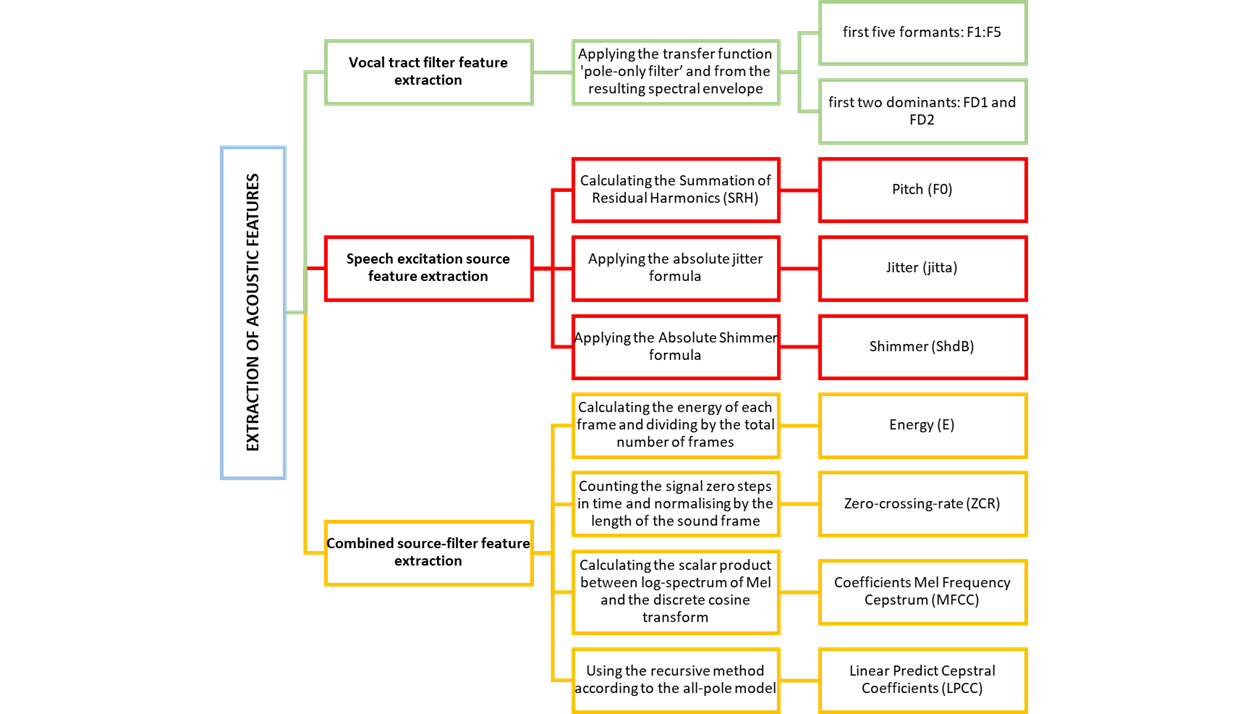}}
	\caption{Diagrammatic representation of the sequence of feature extraction methods used}
	 \label{fig:FE}
\end{figure}

Fig. \ref{fig:FE} shows a diagrammatic representation of the sequence of feature extraction methods used.

Among the features of the vocal tract filter, the first five formant frequencies (F1 to F5), and the first two dominant ones, FD1 and FD2, were extracted. The speech signal was divided into audio frames of 25 ms duration, with a frame shifting of 10 ms. In this way the original non-stationary speech signal is approximated with several shorter almost-stationary frames and Fourier analysis may be applied \cite{Lindasalwa2010}.  

A Hamming window is then applied on each frame in order to minimize the signal discontinuity at the beginning and end of each segment and thus minimize the spectral distortion:

\begin{equation}
\label{eq_2}
  w(n)= 0.54 - 0.46 \phantom{i} cos\left(\frac{2 \pi n}{N-1}\right), \phantom{xxxxx} 0 \le n \le N-1
\end{equation}

where $N$ is the number of samples per frame, so that the signal after the Hamming window is expressed as $y_k(n)=x_k(n) w(n)$ where $n$ refers to the $n$-th sample in the considered $k$-th frame.

After setting the framing of the signal, both the dominants and formants vocal tract features are computed following \cite{Mohanta2020}.
The envelope shape of the spectrum after passing a linear predictive (LP) filter is derived from each vocal frame and describes the features of the resonance frequencies of the vocal tract.

In further detail, the signal is resampled at 10 kHz, and divided into the frames as described above. An LP filter of order $p$ equal to 10 is set to derive the formant frequencies. With these conditions, the frequency response of the filter has a maximum of five peaks corresponding to the frequencies of the first five formant ones, F1 to F5.

\begin{figure}
\centering
	\subfigure[Autistic subject]{\includegraphics[width=0.45\textwidth]{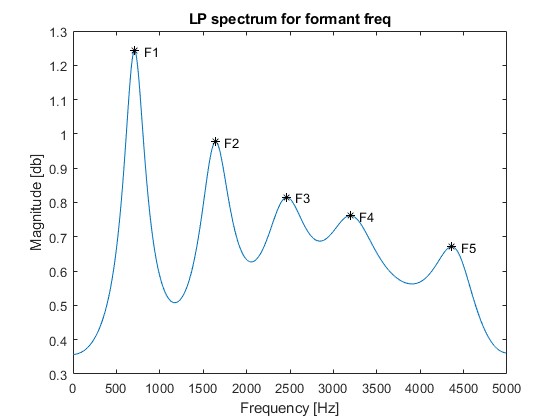}}\label{fig:LP1}
	\subfigure[Typically developing subject]{\includegraphics[width=0.45\textwidth]{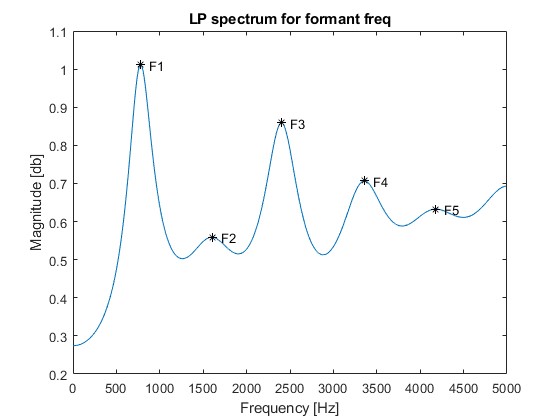}}\label{fig:LP2}
	\caption{LP spectrum showing the formant frequencies obtained from the audio signal of (a) an autistic subject, and (b) a neurotypical one.}
	 \label{fig:LPspectrum}
\end{figure}

Fig. \ref{fig:LPspectrum} shows an example of  an LP spectrum where the first 5 formant frequencies are highlighted.

\begin{figure}
\centering
	\subfigure[Autistic subject]{\includegraphics[width=0.45\textwidth]{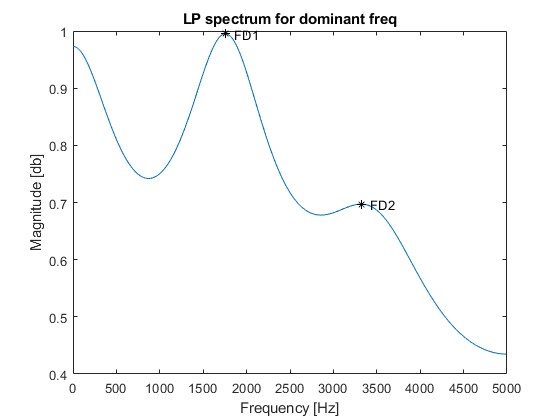}}\label{fig:D1}
	\subfigure[Typically developing subject]{\includegraphics[width=0.45\textwidth]{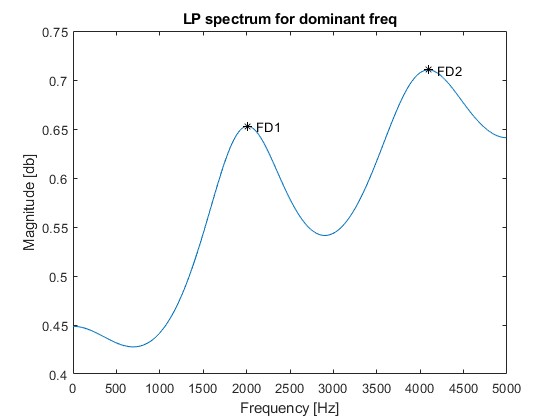}}\label{fig:D2}
	\caption{LP spectrum showing the dominant frequencies obtained from the audio signal of (a) an autistic subject, and (b) a neurotypical one.}
	 \label{fig:Dominant}
\end{figure}

If the filter order $p$ is set equal to 5, the output of the filter shows only 2 peaks, which represent the dominant frequencies FD1 and FD2.
Fig. \ref{fig:Dominant} represents an envelope of the LP spectrum where the dominant frequencies are highlighted. The filter is represented by

\begin{equation}
\label{eq_3}
  H(z)= \frac{1}{1-\sum_{k=1}^{p}\alpha z^{-k}}
\end{equation}

where $p$ is the filter order and $\alpha_k$ are the LP coefficients for $k=1,2,...,p$.

The excitation source features include pitch, closely related to fundamental frequency, shimmer and jitter. 
The method chosen to extract pitch is based on the summation of residual harmonics (SRH) \cite{Drugman2019}. It estimates an auto-regressive (AR) model of the spectral envelope of the speech signal $y(t)$ and derives the residual signal $e_r(t)$ by inverse filtering. For each frame, the amplitude of the spectrum $E_r(f)$ is calculated, which has a relatively flat envelope for speech segments and has peaks at the harmonics of the fundamental frequency $F_0$. From this spectrum, and for each frequency in the range [$F_{0,min}$; $F_{0,max}$], the SRH is calculated as:

\begin{equation}
\label{eq_4}
  SHR(f)= E_r(f) + \sum_{k=2}^{N_{harm}} \left[ E_r(kf) E_r \left(\left(k-\frac{1}{2}\right)f\right) \right]
\end{equation}
where $N_{harm}$ is the number of the first harmonics taken into account, for instance $N_{harm}=5$.
The range of frequencies was set to [70; 400] Hz to account for normal pitch ranges for both men and women. The estimated pitch value $F_0$ for a given frame is the frequency that maximizes $SRH(f)$.



The selected pitch extraction technique guarantees higher performance compared to other methods because of its robustness to interference and environmental noise. 

Once the pitch for each subject's sound segment is estimated, the average $F_0$ is calculated and used to construct the input dataset for the classification methods.

Besides pitch, jitter and shimmer are calculated according to \cite{Drugman2019,Teixeira2013}.
The absolute jitter $Jitt_a$, i. e., the absolute mean difference between consecutive glottal periods, measured in seconds, is defined as:

\begin{equation}
\label{eq_5}
  Jitt_a= \frac{1}{N_F-1}\sum_{i=1}^{N_F-1}|T_i-T_{i-1}|
\end{equation}

in which $T_i$ is the extracted glottal period, corresponding to the instant of time when the maximum of $SRH(f)$ was detected, and $N_F$ is the total number of sound frames.

The absolute shimmer ($ShdB$) is calculated as: 

\begin{equation}
\label{eq_6}
  ShdB=\frac{1}{N_F-1}\sum_{i=1}^{N_F-1} \left|20 \phantom{i} log_{10} \left(\frac{A_{i+1}}{A_i}\right)\right|
\end{equation}

where $A_i$ is the amplitude associated with the pitch value $F_0$ of the $i$-th sound frame and $N_F$ is the total number of sound frames.

Both jitter and shimmer are calculated only between consecutive sound frames.

The combined source-filter features analyzed in this work are energy, zero-crossing
rate, Mel frequency cepstrum coefficient (MFCC) and linear prediction cepstrum coefficient (LPCC).

The energy for each frame is calculated as $E(i)=\frac{\sum_{n=1}^N\left|y_i(n)\right|^2}{N}$, with $i=1,2,...,L$, where $L$ is the total number of frames and $N$ is the frame length. The energy of each frame is normalized to the maximum of its values.

The zero-crossing rate (ZCR) is computed for each frame, evaluating for each sample of the frame if its previous or following sample shows an opposite sign.

\begin{figure}
\centering
	\subfigure[Energy - autistic subject]{\includegraphics[width=0.45\textwidth]{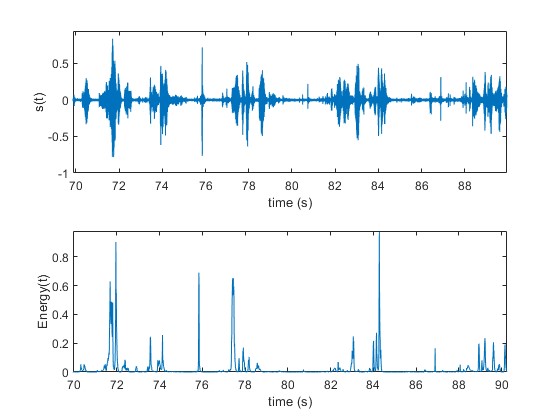}}\label{fig:EZC1}
	\subfigure[Energy -typical development subject]{\includegraphics[width=0.45\textwidth]{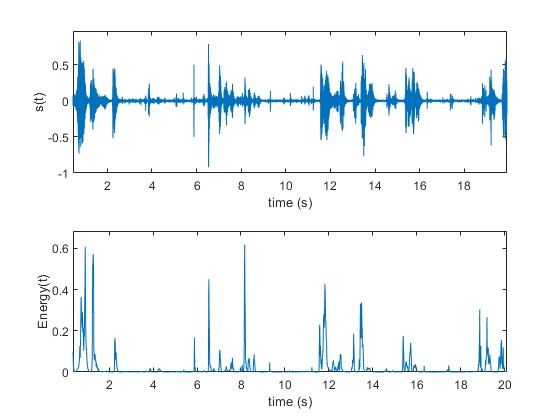}}\label{fig:EZC2}
 \subfigure[Zero-crossing - autistic subject]{\includegraphics[width=0.45\textwidth]{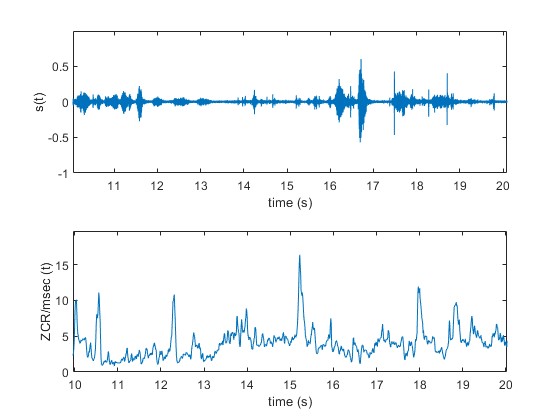}}\label{fig:EZC3}
	\subfigure[Zero-crossing - typical development subject]{\includegraphics[width=0.45\textwidth]{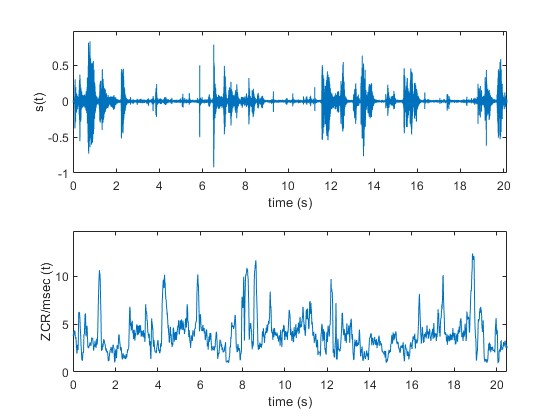}}\label{fig:EZC4}
	\caption{Energy and zero-crossing rate of the audio signal for two different individuals.}
	 \label{fig:Energy-ZeroCrossing}
\end{figure}

\begin{figure}
\centering
	{\includegraphics[width=1\textwidth]{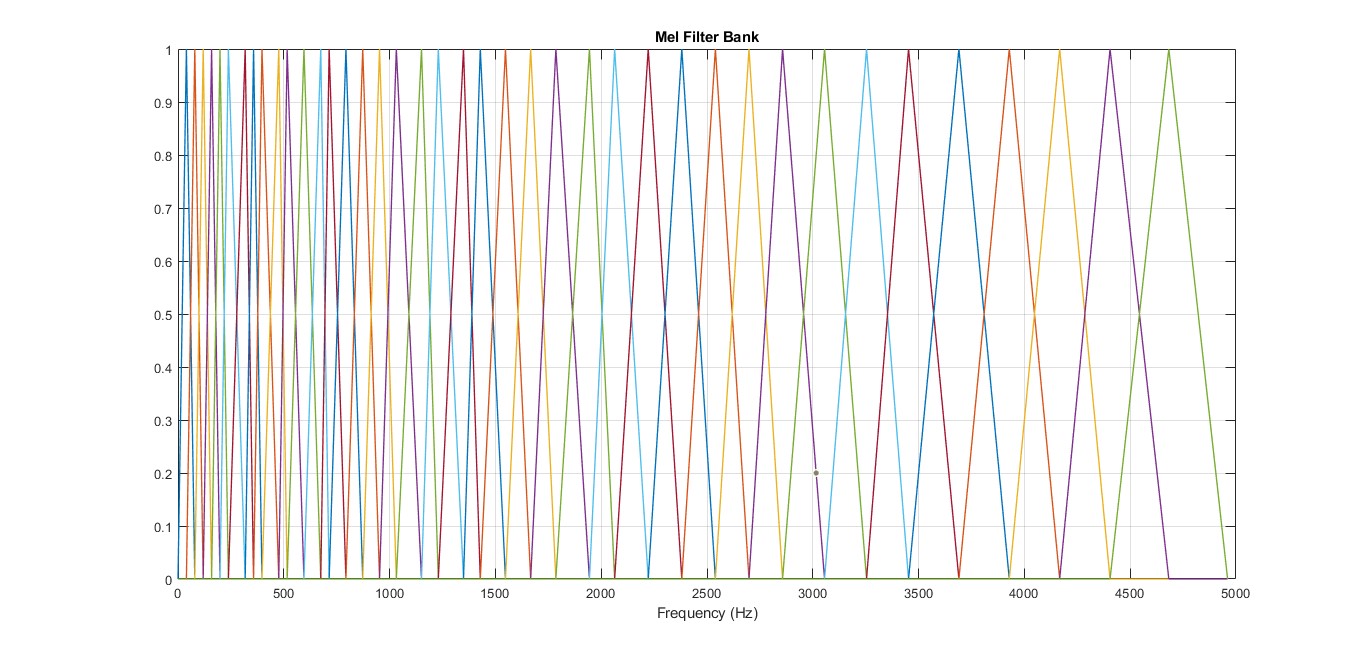}}
	\caption{Frequency response of the Mel filter bank in the range 0-5 KHz with triangular filter shape and M=40.}
	 \label{fig:Filter}
\end{figure}

To calculate  MFCC coefficients, the audio signals were re-sampled at 10 kHz.
Their power is calculated as $P(i)=\sum_{k=0}^{N-1}\left|X_i(k)\right|^2$, where $k$ refers to the $k^{th}$ frequency sample of the spectrum of the $i^{th}$ frame and $X_i(k)=\sum_{n=0}^{N-1}x_i(n)e^{-\frac{2\pi jnk}{N}}$, $k=0,1,...,N-1$. Each $m^{th}$ triangular filter of the Mel filter bank is designed according to \cite{Abdul2022}:

\begin{equation}
H_m(k)= \left\{ \begin{array}{ll}
0, \phantom{xxxxxxxxxxxxxxi} k<f(m-1)\\
\\
\frac{k-f(m-1)}{f(m)-f(m-1)},  \phantom{xxxxxi} f(m-1)\le k<f(m)\\
\\
1,  \phantom{xxxxxxxxxxxxxxi} k=f(m)\\
\\
\frac{f(m+1)-k}{f(m+1)-f(m)},  \phantom{xxxxxi} f(m)<k\le f(m+1)\\
\\
0, \phantom{xxxxxxxxxxxxxxi} k>f(m+1)\\
\end{array} \right.
\label{eq_7}
\end{equation}

where $m=0,1,...,M-1$, $f$ is the normalized discrete frequency, $f(m)$ is the centre frequency of the triangular filter, and $M$ number of triangular filters. $M$ may be chosen between 4 and 160, and we set $M=40$ as usual for speech recognition applications. The resulting filter bank is shown in Fig. \ref{fig:Filter}. 
The figure shows the triangular filters that are linearly spaced from 0 to 1 kHz, and equally spaced in logarithmic scale after 1 KHz, following the typical filtering ability 
of the human ear.

The Mel spectrum is calculated as 

\begin{equation}
\label{eq_8}
  S_{Mel}(m)=\sum_{k=0}^{N-1} \left|X_i(k)\right|^2H_m(k), \phantom{xxxxxxxxx} 0 \le m \le M-1
\end{equation}

The Mel coefficients are calculated as 

\begin{equation}
\label{eq_9}
  c(n)=\sum_{m=0}^{M-1}log_{10}(s_{Mel}(m))\phantom{i}cos\left(\frac{\pi n (m-0.5)}{M}\right)
\end{equation}

where $n=0,1,...,C-1$ with $C$ the total number of MFCC filters.
This operation converts the frequency spectrum into the time domain, keeping only the first C most significant coefficients: the first 12 Mel coefficients are considered here.

LPCC coefficients were calculated through a recursive method that allows to transform LPC filter parameters into the LPC cepstrum according to the all-pole model.
The recursive method is described as

\begin{equation}
c_n= \left\{ \begin{array}{ll}
a_n, \phantom{xxxxxxxxxxxxxxxxxxx} n=1\\
\\
a_n + \sum_{k=1}^{n-1} \frac{k}{n} c_k a_{n-k}, \phantom{xxxxxi} 1 < n \le p\\
\\
 \sum_{k=1}^{n-1} \frac{k}{n} c_k a_{n-k}, \phantom{xxxxxxxxxi} n > p\\
\end{array} \right.
\label{eq_10}
\end{equation}

where $a_1, a_2,...,a_p$ are the coefficients of the LP filter with order $p=12$.

\subsection{Classification Algorithms}
The dataset samples were classified according to the following 36 feature attributes: the fondamental frequency $F_0$, ZCR, the signal energy, the formant frequencies from $F1$ to $F5$, the dominant frequencies $FD1$ and $FD2$, the first 12 Mel-frequency Cepstrum coefficients, the first 12 LPC coefficients, shimmer, jitter.

Firstly, we considered a classification preprocessing phase in order to mage data normalization and select the best feature attributes for the main classification goal, i.e. to distinguish between autistic and non autistic recorded voices.

We focus on supervised classification, taking into accounts the random forest (RF), support vector machine (SVM), logistic regression (LR), and Naive Bayes (NB) algorithms \cite{Demsar2013}.

In the next section, the implemented classification chain is described in detail, starting with a description of the experimental dataset used to provide the results obtained.

\section{Experimental Results}
\label{ER}
In this section, we describe the dataset used for experimental validation of the proposed classification chain.
\subsection{Dataset}

In this work we analyze the speech sequences from ASDBank Dutch Asymmetries Corpus \cite{ASDBank}, and specifically the SK sub-corpus that comprises vocal samples of 46 children with ASD and 38 children with typical development (TD), both groups aged between 6 and 12 years (average 9 years). Most of them were boys and all were native Dutch speakers. These samples were produced between 2007 and 2012 in the framework of a research project at the University of Groningen, and are made available to the scientific community for research purposes \cite{Hendriks2014,Kuijper2015}.

The dataset consists of 84 examples in total, and there are no missing data. Two classes are defined: ASD with 46 subjects out of the total number of examples (54.76\%) and TD with 38 (45.24\%).

\subsection{Data Preparation}
As pointed out earlier, there are no missing data in the dataset. First, all 36 attributes were normalized in the range $[-1,1]$, according to a min-max normalization procedure \cite{Demsar2013}.

It is to be noted that a large number of attributes was defined, disproportionate to the number of available examples, hence classification algorithms may incur the so-called "curse of dimensionality" issue.

Following the rule of thumb that the number of features should be approximately equal to 1/10 of the number of examples that make up the dataset, a number of attributes equal to 8 should be selected. To reduce the number of features to the desired amount, the supervised feature selection strategy used was the T-test-based classification, which assigns a positive score to attributes whose difference in mean values between the two different classes is large.

The classification phase of this work was conducted on Orange, an open source toolkit for data visualization, machine learning and data mining \cite{Demsar2013}. Therefore, at each 5-fold cross-validation cycle, four folds are used as the training set and a ranking is performed on them. From the ranking, only the eight best variables are selected to form the reduced test set, the remaining fold, which represents the input of the classifiers.  The features selected to compose the best set of features were the first eight: 
\begin{itemize}
    \item ZCR;
    \item the fundamental frequency $F_0$;
    \item the MFCC coefficients 3,4,6,8, and 12;
    \item the dominant frequency FD1
\end{itemize}
  
\subsection{Automatic Classification}
As stated above, we employed the following classification techniques: the
RF, SVM, LR, and NB algorithms.
Moreover, we validated the performance of classification algorithms by using the following metrics \cite{Yalug2021}:
\begin{equation}
    Accuracy=\frac{TP+TN}{TP+TN+FP+FN}
\end{equation}
\begin{equation}
    Precision=\frac{TP}{TP+FP}
\end{equation}
\begin{equation}
    Recall=\frac{TP}{TP+FN}
\end{equation}
where $TP$ is the number of true positives, $TN$ the number of true negatives, $FP$ the number of false positives, and $FN$ the false negatives.

A metric that combines the values of both of the above measures is the F measure, which is a harmonic mean between precision and recall:
\begin{equation}
    F_{measure}=\frac{2\cdot precision \cdot recall}{precision+recall}
\end{equation}
\subsection{Experimental Result Evaluation}
\begin{figure}
\centering
	{\includegraphics[width=1\textwidth]{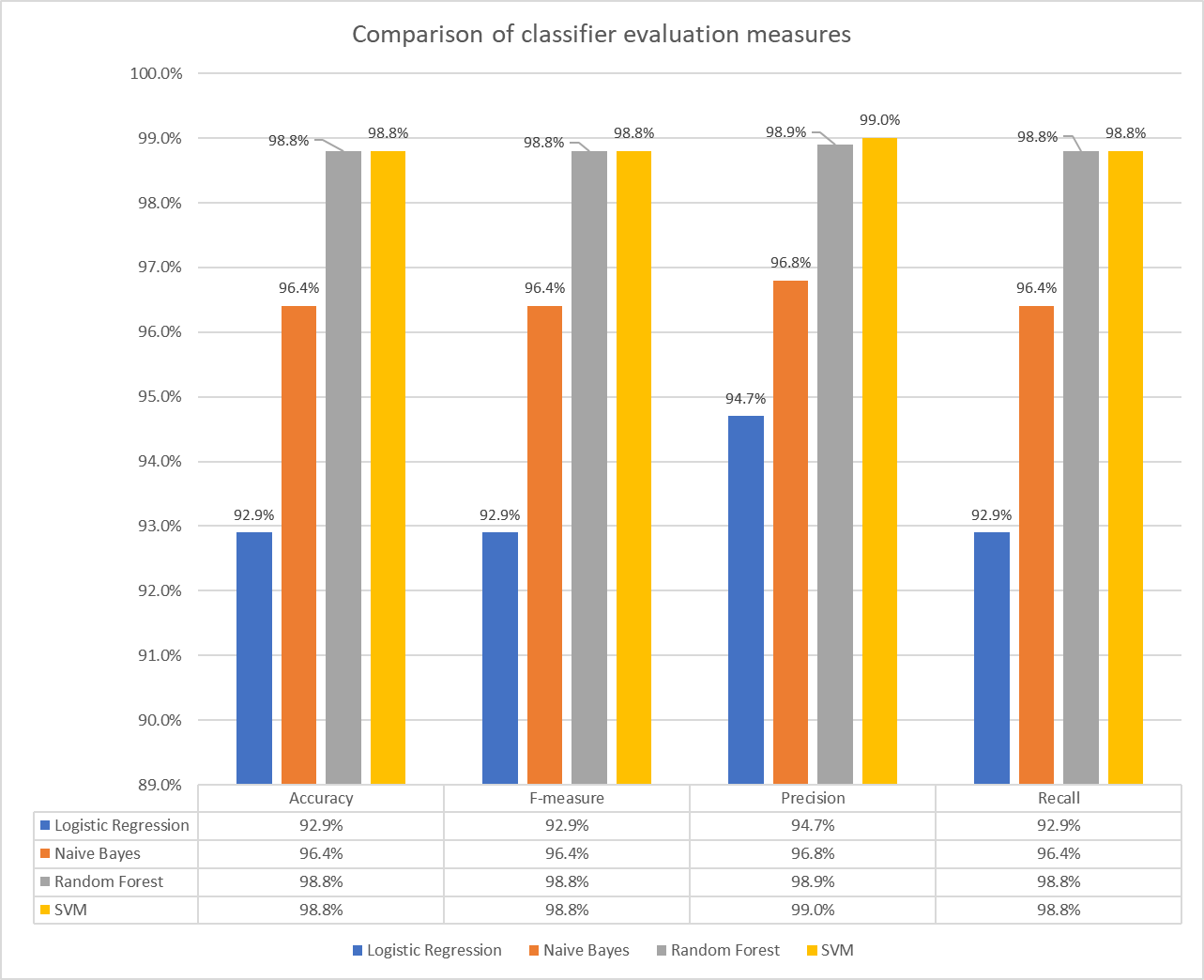}}
	\caption{Classification results.}
	 \label{FIGURA_MODIFICATA}
\end{figure}

Figure \ref{FIGURA_MODIFICATA} shows the classification results. First we highlight the average results derived from the 5-fold cross-validation; at each cross-validation cycle we pinned the average results between ASD and TD classes and at the end of the procedure we extracted the average for Accuracy, $F_{measure}$, Precision and Recall. The best performing classifiers are SVM and RF.

\begin{table}
\centering
\caption{Confidence intervals of the classification algorithms.}\label{tab1}
\begin{tabular}{|l|l|l|l|}
\hline
{\bfseries Classifier} &  {\bfseries Lower limit} & {\bfseries Estimated average accuracy} & {\bfseries Upper limit}\\
\hline
LR &  0.8291 & 0.9294 & 1\\
NB &  0.9326 & 0.9640 & 0.9954\\
RF & 0.9632 & 0.9882 & 1 \\
SVM & 0.9632 & 0.9882 & 1 \\
\hline
\end{tabular}
\label{Estimated average accuracy}
\end{table}
Table \ref{Estimated average accuracy} shows the confidence intervals of the punctual accuracy estimated so far. It turns out that SVM and RF show the same performance even in terms of confidence intervals.

\begin{table}
\centering
\caption{ASD subjects: $F_{measure}$, $Precision$ and $Recall$.}\label{tab2}
\begin{tabular}{|l|l|l|l|}
\hline
{\bfseries Classifier} &  {\bfseries F-measure} & {\bfseries Precision} & {\bfseries Recall}\\
\hline
LR &  0.932 & 0.976 & 0.891\\
NB &  0.967 & 0.978 &0.957\\
RF & 0.989 & 0.979 & 1 \\
SVM & 0.989 & 1 & 0.978\\
\hline
\end{tabular}
\label{PrecisionAndRecall}
\end{table}
However, Table  \ref{PrecisionAndRecall} shows a difference between SVM and RF in terms of precision, which is higher for SVM, although all classifiers score high in precision.

\subsection{Statistical Analysis}
To validate the results obtained in terms of the quality of feature extraction, a parametric-type statistical test, the $t$-test, was performed, to assess the significance value by $p$-value. The starting hypothesis, called the null hypothesis $H_0$, is no difference in the averages of the distribution of attribute values between the "ASD" and "TD" classes; at the beginning of the test, the threshold value $\alpha =0.05$ is then set, which indicates the level of significance with which to compare the $p$-value. If the $p$-value is less than $\alpha$, then we reject the null hypothesis because there is a significant difference within the study. The closer the $p$-value is to 0, the more significant the test becomes.

\begin{table}
\centering
\caption{Statistical Analysis Results.}\label{tab2}
\begin{tabular}{|l|l|l|l|l|}
\hline
{\bfseries Features} &  {\bfseries ASD mean} & {\bfseries TD mean} & {\bfseries p-value} & $H_1$ (5\%)\\
\hline
{\bfseries ZCR} &  0.0987 & 0.167 & $< 0.05$ & Accepted\\
\hline
$\mathbf{F}_0$ &  289.124 & 272.247 & $< 0.05$ & Accepted\\
\hline
{\bfseries MFCC12} &  -2.976 & -3.784 & $< 0.05$ & Accepted\\
\hline
{\bfseries MFCC16} &  -3.0188 & -3.875 & $< 0.05$ & Accepted\\
\hline
{\bfseries MFCC6} &  -3.0188 & -3.875 & $< 0.05$ & Accepted\\
\hline
{\bfseries MFCC3} &  -1.0186 & -3.113 & $< 0.05$ & Accepted\\
\hline
{\bfseries MFCC4} &  -5.0150 & -6.360 & $< 0.05$ & Accepted\\
\hline
{\bfseries MFCC8} &  -3.130 & -3.792 & $< 0.05$ & Accepted\\
\hline
$\mathbf{FD}_1$ &  1853.942 & 1769.413 & $< 0.05$ & Accepted\\
\hline
$\mathbf{F}_1$ &  798.0834 & 820.146 & $0.00956$ & Accepted\\
\hline
$\mathbf{F}_2$ &  1827.998 & 1817.923 & $0.206$ & Rejected\\
\hline
\end{tabular}
\label{t-test}
\end{table}

Table \ref{t-test} shows that the alternative hypothesis $H_1$, according to which with $95\%$ probability the null hypothesis $H_0$ can be rejected, is accepted and it can be confirmed with statistical significance that the differences between the two class means are different from 0. For completeness, we also reported statistics on the first two formant frequencies $F_1$ and $F_2$. It turns out that only the first formant frequency $F_1$ obtains a $p$-value less than $0.05$.

\section{Conclusions}
\label{CC}
Examining autistic prosody is essential because acoustic irregularities might contribute to the socio-communicative changes linked with the disorder. Our work aims to support the diagnosis of autism with more quantitative and objective assessments based on autistic speech.
We attempted to create a clear and defined framework, both for the extraction part, exploiting more complex measures of speech recognition, such as MFCC and LPCC, and for the classification part, defining a systematic cross-validation and filtering of attributes, which is often missing inliterature.
In the future our results will need to be validated by testing the classifiers on a larger number of examples and by generalising the models as much as possible by training and testing them on more languages, different ages or different tasks. Specifically, the focus will be on automatic voice classification of adult autistic individuals.

\end{document}